\documentclass[namedreferences,hyperref,optionalrh]{spr-sola}
\usepackage{graphicx}        
\usepackage{color}           

\chardef\us=`\_

\begin{document}

\begin{frontmatter}
\title{Classifying different types of solar wind plasma with uncertainty estimations using machine learning}

\author[addressref={aff1},corref,email={Thomas.Narock@goucher.edu}]{\inits{T.W.}\fnm{Tom}~\snm{Narock}\orcid{0000-0002-9785-4496}}

\author[addressref=aff2]{\inits{S.P.}\fnm{Sanchita}~\snm{Pal}\orcid{0000-0002-6302-438X}}

\author[addressref=aff1]{\inits{A.A.}\fnm{Aryana}~\snm{Arsham}\orcid{0000-0002-6472-9023}}

\author[addressref={aff3,aff4}]{\inits{A.N.}\fnm{Ayris}~\snm{Narock}\orcid{0000-0001-6746-7455}}

\author[addressref=aff4]{\inits{T.N-C.}\fnm{Teresa}~\snm{Nieves-Chinchilla}\orcid{0000-0003-0565-4890}}

\address[id=aff1]{Department of Computer and Data Science, Goucher College, Baltimore, MD, 21204, USA}
\address[id=aff2]{NASA Postdoctoral Program Fellow, NASA Goddard Space Flight Center, Greenbelt, MD 20771, USA}
\address[id=aff3]{ADNET Systems Inc, Bethesda, MD, 20817, USA}
\address[id=aff4]{Heliophysics Science Division, NASA Goddard Space Flight Center, Greenbelt, MD 20771, USA}

\runningauthor{Narock et al.}
\runningtitle{Classifying Solar Wind With Uncertainty}

\begin{abstract}
Decades of in-situ solar wind measurements have clearly established the variation of solar wind physical parameters. These variable parameters have been used to classify the solar wind magnetized plasma into different types leading to several classification schemes being developed. These classification schemes, while useful for understanding the solar wind’s originating processes at the Sun and early detection of space weather events, have left open questions regarding which physical parameters are most useful for classification and how recent advances in our understanding of solar wind transients impact classification. In this work, we use neural networks trained with different solar wind magnetic and plasma characteristics to automatically classify the solar wind in coronal hole, streamer belt, sector reversal and solar transients such as coronal mass ejections comprised of both magnetic obstacles and sheaths. Furthermore, our work demonstrates how probabilistic neural networks can enhance the classification by including a measure of prediction uncertainty. Our work also provides a ranking of the parameters that lead to an improved classification scheme with $\sim96\%$ accuracy. Our new scheme paves the way for incorporating uncertainty estimates into space weather forecasting with the potential to be implemented on real-time solar wind data.
\end{abstract}
\end{frontmatter}

\section{Introduction}
     \label{sec1} 
The solar wind is a continuous outflow of magnetized plasma emanating from the Sun's outer atmosphere. Different types of solar wind plasma may emanate from the solar surface or develop in the interplanetary medium \citep[e.g.][]{2017SSRv..212.1345C,  2014JGRA..119.4157C, 2013JAdR....4..221L, 2017LRSP...14....5K}. The varying characteristics of this plasma have been used to categorise the solar wind into different types. The most obvious of the solar wind types are the fast and slow solar wind, but more subtle distinctions such as shocked plasma and structures involving various transient events have also been observed using in situ spacecraft observations. \cite{2019JGRA..124.2406B} provide a detailed study examining the varying properties of the solar wind magnetized plasma.
\par
It is generally accepted that solar wind plasma can be categorised in three major types - (1) coronal-hole-origin plasma (CH) sometimes referred to as fast solar wind mostly originating from the interaction of open field lines and low-lying closed coronal loops, (2) streamer-belt-origin plasma (SB) sometimes referred to as slow solar wind originating from edges of coronal holes near streamer belt regions, the interchange reconnection between field lines from closed streamer belt and open magnetic field regions, and sometimes from the tip of the helmet streamers, and (3) plasma originating from solar transients like coronal mass ejections (CMEs). \cite{2015JGRA..120...70X} show evidence for a fourth type named sector-reversal-region (SR) plasma, which is a sub-type of streamer belt plasma. \cite{2020ApJ...889..153R} used a purely statistics based machine learning approach to determine the number of plasma states in the solar wind.  While \cite{2020ApJ...889..153R} agrees there is evidence for three or four major types, their study showed additional evidence that meaningful distinctions in solar wind plasma can be made up to at least eight types. There is even evidence for subdividing types and defining ten types with clearly identifiable refined categories of transients, such as very fast interplanetary CMEs (ICMEs) \cite{2020ApJ...889..153R}.\par 

Which parameters to use in solar wind classification is another debated topic in the literature. For example, an onboard solar wind classification algorithm was part of the Genesis spacecraft \citep{2003SSRv..105..661N, 2003GeoRL..30.8031R}. The automated algorithm required proton temperature, alpha-particle abundance, and the presence of bidirectional streaming suprathermal electrons to classify the solar wind into one of three types. \cite{2015JGRA..120...70X} on the other hand, used proton-specific entropy ($Entropy$), the proton alfvén speed ($V_a$), and the proton temperature ($T_p$) compared with a velocity-dependent expected temperature ($T_{expected}$) to classify solar wind plasma into one of four states -- CH, SB, SR and ICME.  Other classification schemes utilize the ratio of number densities of heavy ion charge states \citep[e.g.][]{2009GeoRL..3614104Z}; yet, most solar wind spacecraft do not carry solar wind ion composition measurement instruments and therefore heavy ion ratios are generally not available in solar wind data sets \citep{2015JGRA..120...70X}. Alternative plasma state classification schemes have been proposed using only commonly available solar wind magnetic field and plasma parameters. \cite{2020E&SS....700997L}, for example, have proposed a machine learning based classification system using eight parameters - seven plasma characterising parameters and the magnetic field magnitude. \par

\cite{2015JGRA..120...70X} and many related studies \citep[e.g.][]{2017JGRA..12210910C,2020ApJ...889..153R} mainly focus on magnetic clouds \citep[MCs;][]{2005AnGeo..23.2687L} to define the ICME-type of solar wind plasma, which they refer to simply as `ejecta'. MCs are a subset of ICMEs containing an enhanced large-rotating magnetic field vector $\mathbf{B}$ suggestive of a flux rope field configuration, low $T_p$ and low plasma $\beta$ (i.e. the ratio of gas pressure to the magnetic pressure). The definition of MCs is based on earlier work by \cite{1982JGR....87..613K}. In recent years, \cite{2018SoPh..293...25N} have found the magnetic cloud definition to be too restrictive. They introduced the term `magnetic obstacle' (MO) to signify more general magnetic configurations within solar wind transients. MOs generally follow the properties of ICMEs described by \cite{2006SoPh..239..393J} and are limited to cases that have low plasma beta. 
\par

We first consider the simpler approach of \cite{2015JGRA..120...70X} that classifies solar wind plasma in four types - CH, SB, SR and ejecta using three parameters – $Entropy$, $V_a$, and $T_{ratio} = (T_p/T_{exp})$ - which we refer to as the 3-parameter scheme. These three quantities are easily computed from commonly available solar wind parameters including magnetic field magnitude, proton density, temperature, and speed. We are interested in assessing the utility of the 3-parameter scheme when the `ejecta' category as considered by \cite{2015JGRA..120...70X} is generalised to MOs.  We use this as a starting point, noting that the number of solar wind plasma states - and the quantities needed to determine them - is still being debated. We use machine learning models trained with the 3-parameter scheme to classify the solar wind plasma as CH, SB, SR or MO. In this way, we assess whether MOs have plasma states similar to the idealized definition of what \cite{2015JGRA..120...70X} call ejecta. Later, we expand our study to classify another solar wind plasma type that is the compressed regions called sheaths (SHs) that generally precede the MOs or MCs and mostly form during an ICMEs' interplanetary propagation \citep{2017LRSP...14....5K}. To the best of our knowledge, no current plasma type classification scheme looks at the complex sheath region. The SHs are distinctly different from the ejecta definition of \cite{2015JGRA..120...70X}. They have larger $\mathbf{B}$ variations, much higher $T_p$ and plasma density. They are more turbulent than ejectas \citep{2017LRSP...14....5K}. We attempt to classify the SH as a distinct plasma type. The properties of the SH can depend on the type of solar wind plasma preceding it, i.e. the presence of a shock, fast solar wind, or prior ejecta. Thus, a single SH category is likely not a sufficient long term strategy. However, given the lack of research on SH classification we focus here on the SH as a single category. We save for future research the machine learning workflow augmentations needed to identify sub-types of SH.   
\par

Additionally, we explore the effectiveness of using more than three parameters to train our machine learning models and we introduce a measure of prediction uncertainty based on advances in probabilistic machine learning. We are interested in how knowledge of different plasma types can contribute to space weather assessments. For example, \cite{2020SoPh..295..131D}, \cite{2022FrASS...938442N}, and \cite{2024ApJ...972...94P} used deep neural networks to identify MOs in the solar wind relying solely on magnetic field data and the presence of magnetic flux ropes. Yet, \cite{2018SoPh..293...25N} show that not all MOs exhibit a rotation in their magnetic fields indicative of flux ropes. Thus, involving plasma information in MO identification can potentially improve previously established solar wind auto-identification models that focus only on magnetic parameters and rotations of the magnetic field. The machine learning classification studies by \cite{2019ApJ...874..145N} and \cite{2022SpWea..2003149R} utilize both magnetic field and plasma parameters. However, those studies are window-based requiring a multidimensional time series as input. Our study, by contrast takes a multidimensional time point as input to a probabilistic neural network model. Classifying the solar wind point by point instead of over time intervals can provide additional benefits in forecasting and reduced training time. \par
Our manuscript is organised as follows. In Section \ref{sec2} we discuss the data and their preprocessing to derive the solar wind parameters that we further use in this study. Section \ref{sec3} discusses the machine learning model and the process of uncertainty estimation. Section \ref{sec4} presents the results obtained from our model and discusses the evaluation of the results, and Section \ref{sec5} summarises the study and presents our conclusions. An Appendix summarizing the acronyms used throughout is available at the end of the manuscript following Section \ref{sec5}.

\section{Dataset}\label{sec2}
We used one minute resolution solar wind magnetic field and plasma data observed in situ by the Magnetic Field Instrument \citep[MFI;][]{1995SSRv...71..207L} and the Solar Wind Experiment \citep[SWE;][]{1995SSRv...71...55O} onboard the Wind spacecraft. We used these data obtained at 1 au to compute different magnetic and plasma properties that are then used as inputs to our models. The parameters are $Entropy$, $V_a$, $T_{ratio}$ which are considered in the 3-parameter scheme of \cite{2015JGRA..120...70X}, solar wind turbulence properties -- cross helicity $\sigma_c$, residual energy $\sigma_r$ -- that can serve as proxies for solar wind heavy ion composition \citep{2020ApJ...889..153R}, the ratio of gas pressure to the magnetic pressure -- plasma $\beta$ also included in the study by \cite{2020E&SS....700997L}, fluctuations of the magnetic field strength $B_{rms}$ that is an important distinguishing characteristic when studying SHs \citep{2021ApJ...921...57S} and the total pressure ($P_{tot}$) found  useful in effectively distinguishing ICMEs from other solar wind structures \citep{2006SoPh..239..393J}. The definition of each parameters are given below:

\begin{equation}\label{eq1}
    V_a=\frac{B}{\sqrt{\mu_0 m_p N_p}}
\end{equation}
\begin{equation}\label{eq2}
    T_{ratio}=\frac{T_{expected}}{T_p}
\end{equation}

\begin{equation} \label{eq3}
    Entropy=\frac{T_p}{N_p^{2/3}}
\end{equation}
\begin{equation} \label{eq4}
    \sigma_c=\frac{E_+ - E_-}{E_+ + E_-},
\end{equation}
where $E_{\pm}$ represents the trace spectral densities of Elsasser variables $z^{\pm}=V_{sw}\pm V_a$. The fluctuation in $z^{+}$ and $z^{-}$ represents Alfvén wave propagating parallel and anti-parallel to the background magnetic field.
\begin{equation} \label{eq5}
    \sigma_r=\frac{E_v - E_b}{E_v + E_b},
\end{equation}
where the trace spectral density of $V_{sw}$ and $V_a$ is represented by $E_v$ and $E_b$, respectively.
\begin{equation}\label{eq6}
    B_{rms}=\sqrt{\sum_{i=1}^{n}\frac{(B_i-<B>)^{2}}{n}},
\end{equation}
where $<B>$ represents the temporal average of the magnetic field strength $B$ for 10 minute intervals. Equation \ref{eq6} is Equation 4 from \cite{2021ApJ...921...57S}.
\begin{equation}\label{eq7}
    \beta=\frac{N_p k T_p}{B^2/2\mu_0},
\end{equation}
where $k$ is the Boltzman constant and 
\begin{equation}\label{eq8}
    P_{tot}=\frac{B^2}{2\mu_0} +N_p k T_p.
\end{equation}
Where $B$ is the magnetic field magnitude, $m_p$ is the proton mass, $N_p$ is the proton density, $T_p$ is the proton temperature, $T_{expected}$ is the expected temperature defined as a function of solar wind speed ($V_{sw}$) as $\frac{V_{sw}}{258}^{3.113}$ \citep{2015JGRA..120...70X}, and $N_p$ is the solar wind density. $V_a$ is the Alfvén speed in units of km/s and entropy is in units of $eV.cm^2$. \par

The MOs and SHs are collected from  
\href{https://wind.nasa.gov/ICME_catalog/ICME_catalog_viewer.php}{ICME Catalog}. This catalog includes lists of manually identified ICME and MO start and end times. Our considered SHs are the region between the start of the ICME and the start of the MOs. In this catalog, the MOs are divided into five types: Flux-rope (Fr), Small rotation flux-rope (F-), Large rotation flux-rope (F+), Complex (Cx), and Ejecta (E) (note that here E has a different definition from ejecta defined in \cite{2015JGRA..120...70X}) \citep{2018SoPh..293...25N}. Events without evident rotation (E) and those with more than one magnetic field rotation (Cx) were withheld from our machine learning training dataset. The rationale is to train on Fr, F+, and F- plasma characteristics and then evaluate if Cx and E events have similar characteristics. We imposed a 90\% data availability criterion to avoid events with large data gaps. In the end, 308 events of Fr, F-, and F+ type from the \href{https://wind.nasa.gov/ICME_catalog/ICME_catalog_viewer.php}{ICME Catalog} were used as our candidate MOs. 
\par
We then randomly sampled from solar minimum (years 2020 through 2022) and solar maximum (years 2012 through 2014) of OMNI one-minute data \citep{doiSPASEResource} collected at 1 au to obtain a comparable dataset of non-transient solar wind plasma types. Each data point of our OMNI sample was classified as CH, SB, or SR following the categorisation algorithm explained in Section 2.2 of \cite{2015JGRA..120...70X}. Any OMNI points classified as 'ejecta' were discarded. Next, we averaged Wind and OMNI data to 10 minutes to remove small-scale fluctuations. This resulted in a dataset of 12,239 CH, 12,255 SR, 12,248 SB, 12,239 MO, and 12,238 SH points. We note that our approach classifies the plasma state of each point in a time series. In other words, the inputs are the plasma characteristics of a single time point and the output is a classification of solar wind plasma types. We classify the plasma state at each time step and do not evaluate the time series as a whole. As such, we do not look for temporal changes such as rotations of the magnetic field. Instead, the aim of this work is to detect the plasma type at any given time point. \par 

We normalized our data using minimum and maximum values from the entire dataset (maximum of solar cycle 24 and the most recent minimum) as required by neural networks. Data are then randomly split with the standard 80\% for training the neural networks and the remaining 20\% used for evaluating the trained networks.


\section{Uncertainty Estimation}\label{sec3}
This study utilises neural networks (NNs) consisting of an input layer, two hidden layers with 16 and 8 nodes, respectively, and a categorical distribution output layer, which is a discrete probability distribution of a random variable that can take one of $K$ possible categories. 
Our output categorical distribution is implemented using the Tensorflow Probability software library \citep{2017arXiv171110604D}. The other layers in the NNs are standard Dense layers from the Keras library \citep{2018ascl.soft06022C}. \par

Placing a probability distribution over the NN output layer is method of capturing aleatoric uncertainty, which is uncertainty inherent in the observations. Aleatoric uncertainty is one of two main types of uncertainty that one can capture \citep{2017arXiv170304977K}. The other being epistemic uncertainty, which accounts for uncertainty in the model parameters. In a machine learning context, epistemic uncertainty captures our ignorance surrounding the optimal NN weights. Epistemic uncertainty can be reduced given enough data, as more training data leads to better refinement of model parameters. Aleatoric uncertainty, on the other hand, is irreducible. \par 

Most traditional deep learning classification models use a softmax output layer. The softmax function transforms the output of the network into a normalized vector, which are often interpreted as probabilities. For example, given a binary classification problem, the softmax output of [0.8, 0.2] is interpreted as a 80\% prediction for the first category. However, softmax outputs are not true probability distributions and can report high confidence for incorrect predictions when treated as such \citep{2018arXiv180601768S}. \par

We quantify the aleatoric uncertainty using Shannon entropy, which is common for classification problems \citep{2019arXiv191009457H}. If $p_i$ denotes the probability that an observed event belongs to the plasma type $i$, $i$=1...K, the Shannon entropy is defined as $H = -\sum_i p_i ln( p_i )$. Entropy in information theory is analogous to entropy in statistical thermodynamics. Larger values of entropy, $H$, occur when probabilities are close to being equal. For example, when each state has equal probability of occurring then, $p_i= \frac{1}{K}$, for $i=1...K$. This results in the maximum possible Shannon entropy $H = -ln\frac{1}{k}$. In contrast, if the NN assigns a high probability to a particular class, close to one, the entropy will become close to 0. Therefore, $H$ ranges from $(0, -ln\frac{1}{k}]$. Larger values of $H$ correspond to the NN being more uncertain about a particular prediction. In contrast, lower $H$ occurs when an event has a higher probability of belonging to a particular type of solar wind plasma. We will refer $H$ as `uncertainty' from this point on.\par

Techniques for quantifying both aleatoric and epistemic uncertainty do exist \citep{2017arXiv170304977K}. However, there is still much debate surrounding approaches to quantify epistemic uncertainty in neural networks. We leave epistemic uncertainty for future research in which we will investigate how each of the various approaches impacts epistemic uncertainty quantification. \par 

In this work, a user-specified threshold for uncertainty is the maximum allowable $H$ for a prediction. Our model outputs the probability of a solar wind time step belonging to each plasma type and the associated $H$. If the event has an uncertainty, $H$ value, lower than the user-specified threshold the event is classified to the plasma type with the highest probability. In contrast, if the event has an uncertainty greater than or equal to the threshold, it is unclassified, as the uncertainty pertaining to that prediction exceeds the maximum threshold. Using a lower threshold for uncertainty results in fewer predictions, i.e. higher number of events will not be classified. However, the predicted events will have higher certainty. A higher threshold value results in more events being predicted, but the predicted events may have lower accuracy. Therefore, a risk-averse application can select a lower threshold value to ensure that predictions have lower uncertainty. \par

Previous work by \cite{2017JGRA..12210910C} introduced the notion of uncertainty in machine learning and its application to solar wind plasma type classification. That work leveraged a Gaussian Process method and provided a means of classifying an event as `undecided'. We highlight the work of \cite{2017JGRA..12210910C} as being one of the first to quantify uncertainty in space physics context, but note the differences in approach and means of uncertainty quantification. In this work, we explore an alternative means of quantifying and utilizing prediction uncertainty.

\section{Results and Discussion} \label{sec4}
In this Section, we provide the results obtained from our model trained with different sets of parameters to detect different types of solar wind plasma. \par
\subsection{Plasma state auto-identifications using 3-parameter scheme}
\begin{figure}    
\centerline{\includegraphics[width=1.0\textwidth,clip=]{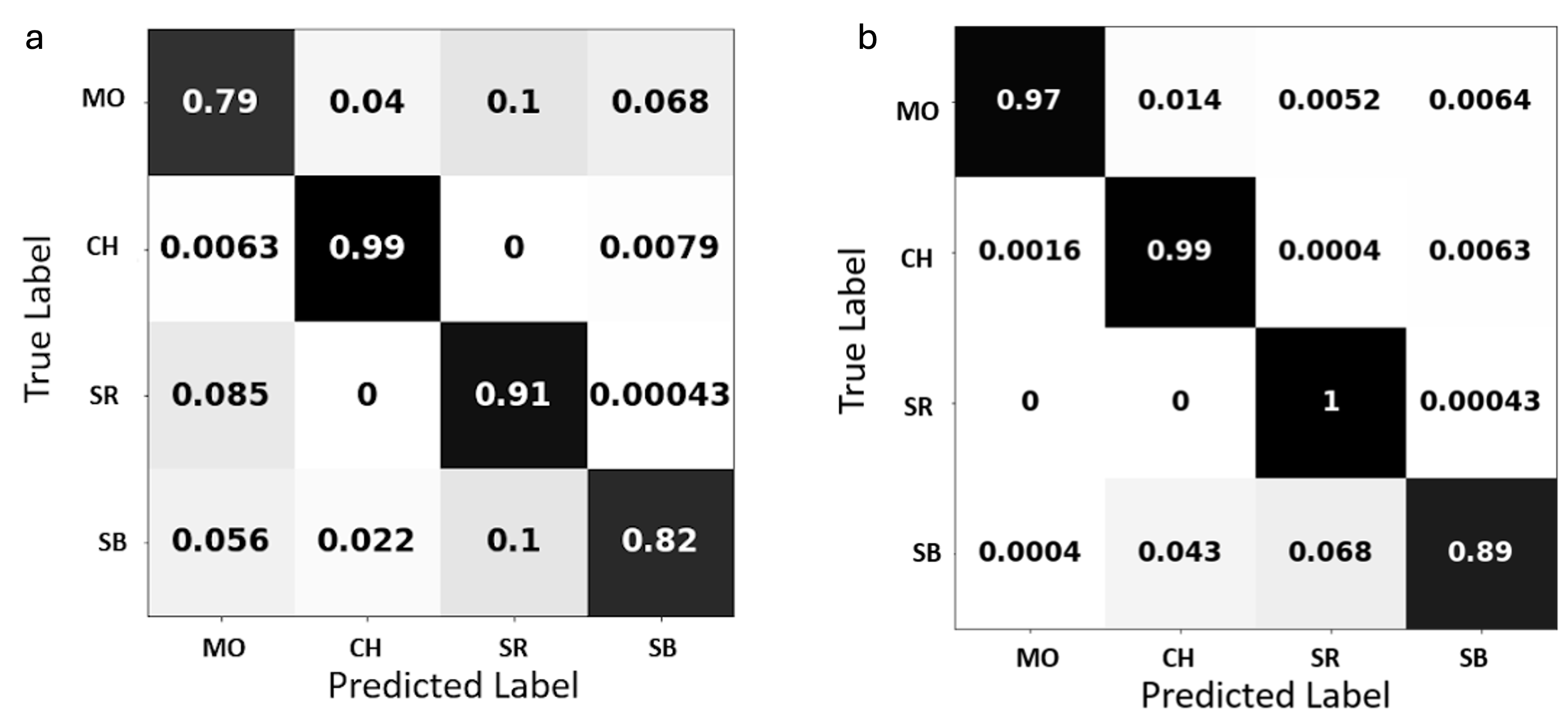}}
\caption{Confusion matrix derived while evaluating (a) 3PNN model and (b) 8PNN model to classify magnetic obstacle (MO), coronal-hole-origin plasma (CH), streamer-belt plasma (SB), sector-reversal-region plasma (SR).
}
\label{fig1}
\end{figure}

In our first experiment, we tested the 3-parameter scheme’s ($V_a$, $T_{ratio}$, $Entropy$) ability to identify MOs using a probabilistic NN hereafter called 3PNN. This experiment categorises solar wind plasma into four ($K=4$) different types: CH, SR, SB and MO. The output of our model is a probability distribution indicating the likelihood of each plasma type. The probability values were used to compute $H$ as a measure of uncertainty. Figure \ref{fig1} and Table \ref{t1} show the model evaluation results. All predictions were used in generating Figure \ref{fig1} and Table \ref{t1}, i.e. no uncertainty filtering was applied. Figure \ref{fig1}(a) indicates high model accuracy for CH and SR, but lower accuracy for MO and SB. Additional summary statistics are given in Table \ref{t1}, which are based on the number of True Positive (TP), False Negative (FN), and False Positive (FP) predictions. These values are then combined to find: $Recall = \frac{TP}{TP + FN}$ , $Precision = \frac{TP}{TP + FP}$ , and $F1 score$, which is the harmonic mean of $Precision$ and $recall: F1 = \frac{2TP}{2TP + FP+FN}$. We mention these metrics in the first column of Table \ref{t1}. We provide average $precision, recall$, and $F1 values$ across all categories when summarizing our results in tables. This is the so-called macro average, which is computed by taking the arithmetic mean (unweighted) across the categories. This approach treats each category with equal significance.\par

\begin{table} \
\caption{ Macro results showing the performance of 3PNN and 8PNN models in identifying CH, SR, SB and MO.
}
\label{t1}
\begin{tabular}{ccc}     
\hline                     
Metrics & Values for 3PNN & Values for 8PNN\\
\hline
Macro $F1$ & 0.8764&0.9628 \\
Macro $Precision$   &0.8783&0.9643  \\
Macro $Recall$ & 0.8779&0.9636\\
$Accuracy$ & 87.79\%&96.32\%\\
\hline
\end{tabular}
\end{table}

\begin{figure}    
\centerline{\includegraphics[width=1.0\textwidth,clip=]{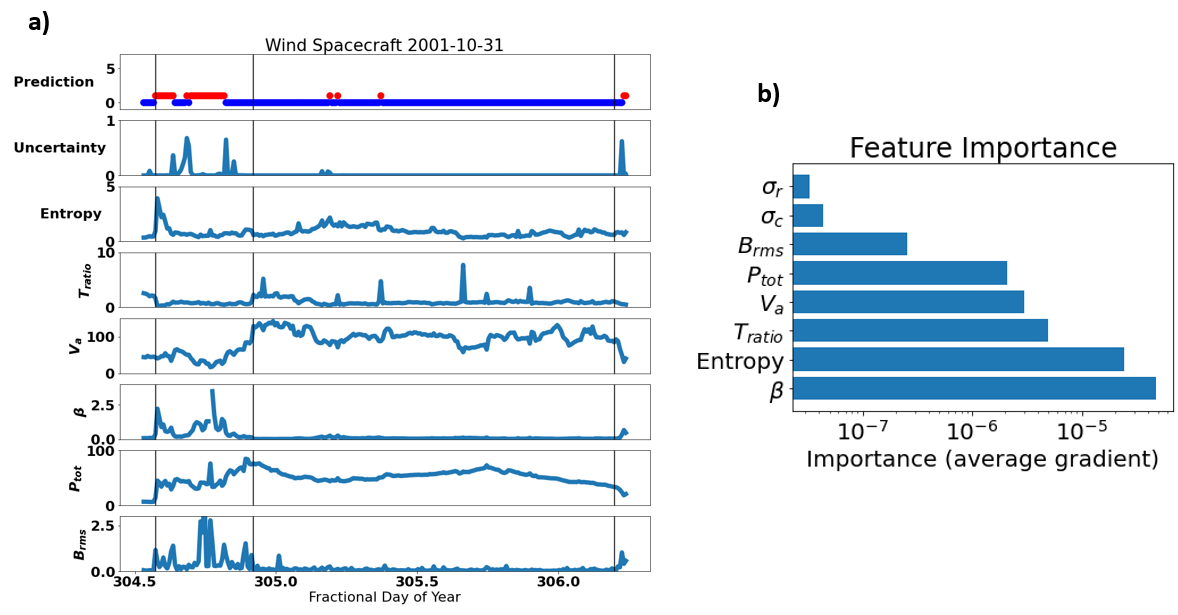}}
\caption{(a) Example of plasma state classification during 2021/10/31 13:00 - 2021/11/02 05:00 using 3PNN model. The top panel shows the classification with blue indicating MO and red indicating one of the other states. Prediction uncertainty, $Entropy$, $T_{ratio}$, $V_a$, $\beta$, $P_{tot}$ and $B_{rms}$ are shown from the top to bottom, respectively. (b) Feature importance analysis for the 8PNN model input. }
\label{fig2}
\end{figure}

Figure \ref{fig2}(a) shows the 3PNN model output applied to an event from the \href{https://wind.nasa.gov/ICME_catalog/ICME_catalog_viewer.php}{ICME Catalog} during October, 2001. The top panel of the figure shows the plasma state prediction with blue indicating magnetic obstacle and red indicating one of the other states. Vertical black lines indicate the start of an SH, an MO, and the end of an MO as determined by human experts in the catalog. At this stage, the SH is not considered as a separate plasma state. Therefore, we see our model output alternates between MO and other types during the SH. Results also indicate that our model’s prediction uncertainty is highest during this SH period (Figure \ref{fig2}(a) 2nd panel). Uncertainty thresholding will be explored in later Sections. We note that this event is preceded by a shock, which is likely leading to the inaccurate prediction of plasma just before the SH. Shocks are known to impact the generally observed relationships among the three parameters of $V_a$, $T_{ratio}$, and $Entropy$ in the 3PNN model \citep{2015JGRA..120...70X}.

\subsection{Plasma state auto-identifications using 8-parameter scheme}
The 3-parameter classification scheme of \cite{2015JGRA..120...70X} is a good place to start given its utilization of readily available parameters and the ease of their calculation. However, as noted in Section \ref{sec1}, other studies have proposed additional parameters for classifying solar wind plasma type. We next expanded the number of inputs to the probabilistic NN model to include five additional parameters: $\sigma_c$, $\sigma_r$, $B_{rms}$, plasma $\beta$, and $P_{tot}$. This resulted in an 8-parameter classification scheme. Figure \ref{fig1}(b) shows the confusion matrix resulting from this 8-parameter neural network hereafter called 8PNN. Overall summary statistics for this model are given in the third column of Table \ref{t1}. \par

Immediately evident when comparing the performance of 3PNN and 8PNN models provided in Figure \ref{fig1} and Table \ref{t1} is the increase in prediction accuracy from additional inputs. Yet, neural networks are black-boxes in which it is difficult to assess how the inputs are leading to the outputs and which inputs are most important. A commonly used techniques for quantifying input feature importance is Integrated Gradients (IG). The IG technique is a gradient-based explanation method that tries to explain a given prediction by using the gradient of the output with respect to the input features. Features with larger gradient magnitudes have a greater impact on the output and are considered more important. The IG results for the inputs of the 8PNN model are shown in Figure \ref{fig2}(b). From Figure \ref{fig2}(b) it is evident that despite $\sigma_c$ and $\sigma_r$ being proxies for the $O^{7+}/O^{6+}$ ratio \citep{2020ApJ...889..153R}, our model finds them of limited importance in our plasma type classifications. We find that the addition of $P_{tot}$ and plasma $\beta$ are important in classification of solar wind plasma type.\par

The confusion matrix provides the proportion of correct and incorrect answers. However, it does not tell us where within a MO interval predictions are incorrect. To explore this, we divided each MO into thirds and calculated the number of incorrect predictions in each third. We found that 2\% of the first third, 1\% of the middle third, and 5\% of the final third predictions were incorrect. We hypothesize that the final third of the event has higher model inaccuracy as this is the trailing edge of the magnetic obstacle and there is not a clean transition from magnetic obstacle to other solar wind plasma types. This notion is supported by research on the presence of open field lines at MO trailing edges. A reconnection at MO front boundaries with the surrounding field lines may leave piled-up reconnected open field lines at the MO trailing edges \citep{2015JGRA..120...43R, 2020GeoRL..4786372P, 2022FrASS...9.3676P, 2022AdSpR..70.1601P}. This may peel off an MO's twisted outer layers \citep{2021A&A...650A.176P}, significantly impact the conservation of magnetic flux and helicity inside ICME flux ropes \citep{GOPALSWAMY201835, 2017ApJ...851..123P} and thus erode them. 

\subsection{Identifying sheath as a separate plasma state}
\begin{figure}    
\centerline{\includegraphics[width=1.0\textwidth,clip=]{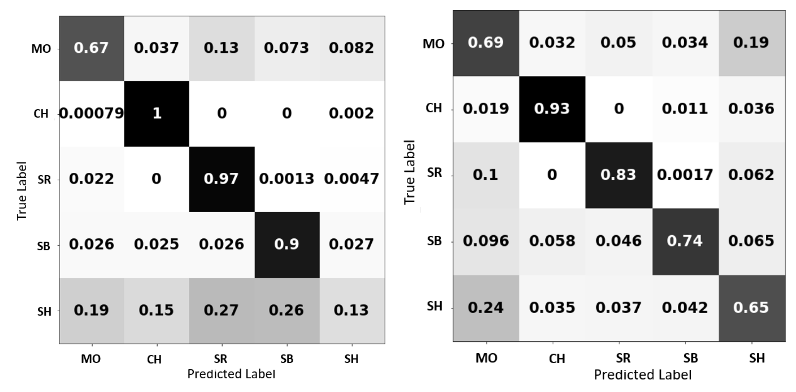}}
\caption{ Confusion matrix for (a) 3PNN and (b)8PNN models for the classification of magnetic obstacle (MO), coronal-hole-origin plasma (CH), streamer-belt plasma (SB), sector-reversal-region plasma (SR), and sheath (SH).
}
\label{fig3}
\end{figure}

The 3PNN and 8PNN models discussed above do not consider SHs that sometimes precede MOs as a separate plasma type. Identification of SHs is beneficial for space weather assessment as it can provide advanced notice of MOs and imminent geoeffective structures given that SH regions often preceed MOs and have been shown to sometimes contain structures capable of causing geomagnetic storms \citep{2017LRSP...14....5K}. To accommodate SH identification, we train our 3PNN and 8PNN models with five output categories ($K=5$). The fact that SHs are known to be highly variable is reflected in the confusion matrices shown in Figure \ref{fig3}(a) and (b) for the 3PNN and 8PNN, respectively.

From Figure \ref{fig3} and Table \ref{t2} it is evident that neither approach adequately distinguishes SHs. While the 8PNN does a better job compared to the 3PNN for SH identification, it still regularly misclassified SHs as MOs and vice versa. Our models' limited success in uniquely identifying SH seems to confirm recent research showing the leading edges and SHs of ICMEs to be chaotic and vary significantly from event to event \citep{2017LRSP...14....5K, 2020ApJ...904..177S, 2022A&A...665A..70T}. Interestingly, when more input features are used, our model is better able to distinguish the SH from CH, SR, and SB plasma; yet, this model still confuses SH and MO plasma states. \par

\begin{table} 
\caption{ Macro results for 3PNN and 8PNN in classification of plasma in MO, CO, SB, SR, and SH.}
\label{t2}
\begin{tabular}{ccc}     
\hline                     
Metrics & Values for 3PNN & Values for 8PNN \\
\hline
Macro $F1$ & 0.601 & 0.771 \\
Macro $Precision$ & 0.600 & 0.778 \\
Macro $Recall$ & 0.604 & 0.769 \\
$Accuracy$ & 60.53\% & 76.94\% \\
\hline
\end{tabular}
\end{table}

\begin{figure}    
\centerline{\includegraphics[width=1.0\textwidth,clip=]{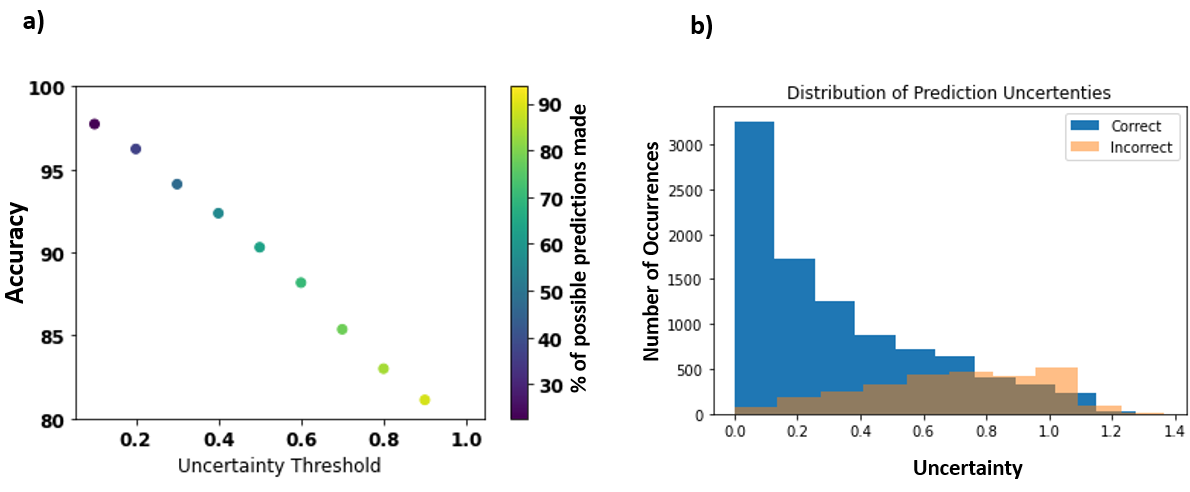}}
\caption{(a) Prediction accuracy of 8PNN model as a function of user-specified uncertainty threshold. Lowering the threshold leads to a trade-off between making fewer predictions, but with increased accuracy. (b) Uncertainty values for correct and incorrect predictions of the 8PNN model. Incorrect predictions are generally associated with larger uncertainty values. Correct predictions are frequently associated with lower uncertainty values; however, there are a number of predictions that were correct but the algorithm still had a high uncertainty about its choice.}
\label{fig4}
\end{figure}

The primary benefit of using probabilistic NNs is their ability to quantify prediction uncertainty. Our model captures the aleatoric uncertainty, which is due to irreducible noise in the data and results from the stochastic nature of solar wind processes generating the data. We use this prediction uncertainty (discussed in Section \ref{sec3}) to assess if we should trust a prediction. Predictions with an uncertainty above a user-defined threshold are not trusted and reclassified as `no prediction'. Figure \ref{fig4}(a) illustrates how $Accuracy$ changes as this user-defined threshold is varied in the 8PNN model. Larger uncertainty threshold values in Figure \ref{fig4}(a) indicate the user is willing to accept predictions the model that has little confidence in. The vertical color bar in Figure \ref{fig4}(a) shows the percentage of predictions that are kept at that threshold. For example, with a threshold value of 0.75 the model is confident making predictions on $\sim~70\%$ of the data resulting in an overall prediction accuracy of 85\%. Uncertainty estimation helps avoid predicting when the model is unsure. Typically, incorrect predictions are associated with higher uncertainty values. However, as evident in Figure \ref{fig4}(b), this is not always the case. All the test data were passed through the 8PNN model. The resulting classification predictions were evaluated for correctness and their prediction uncertainty was recorded. Figure \ref{fig4}(b) shows the distributions of these correct and incorrect predictions. In some cases, the network is highly uncertain, but correct. In a few cases, the network is highly certain, but produces wrong classifications. \par

\begin{figure}    
\centerline{\includegraphics[width=1.0\textwidth,clip=]{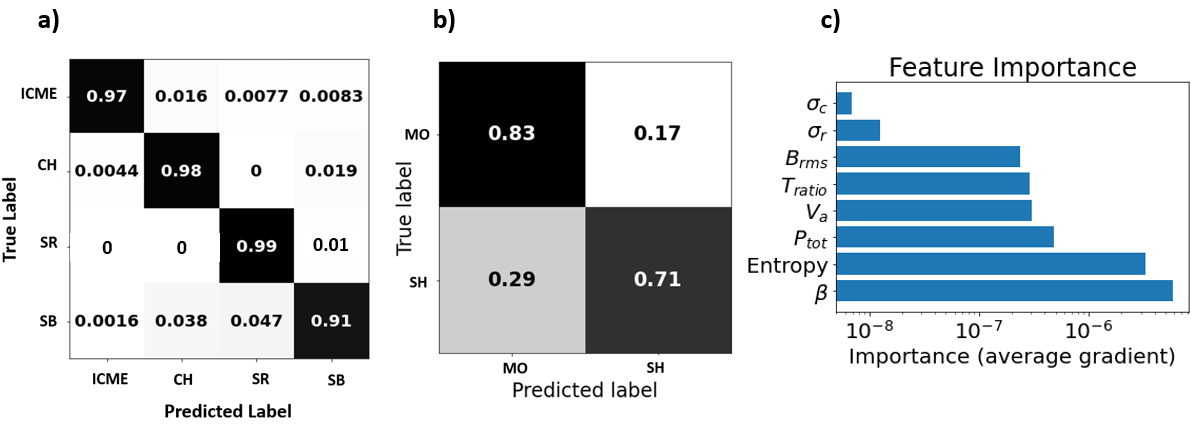}}
\caption{(a) Confusion matrix for the 8PNN model for the classification of ICME, CH, SB, and SR. (b) Confusion matrix of 8PNN classifying ICMEs as MOs and SHs. (c) Feature importance analysis for distinguishing between SHs and MOs.}
\label{fig5}
\end{figure}

We next combined MO and SH into a single category to account for the NNs inability to distinguish the two. This makes sense physically given that SHs often precede MOs and both can be considered as part of ICME structures. We indicate this category as ICME. We then re-trained our 8PNN model to identify solar wind plasma as CH, SR, SB or ICME. Figure \ref{fig5}(a) shows the confusion matrix of the model and Column 2 of the Table \ref{t3} shows the corresponding macro metrics.

\begin{table} \
\caption{ Macro results showing the performance of 8PNN in identifying (a) CH, SR, SB and ICME and subsequently distinguishing ICME values into (b) MO and SH.
}
\label{t3}
\begin{tabular}{ccc}     
\hline                     
Metrics & 8PNN(a)& 8PNN(b)\\
\hline
Macro $F1$ & 0.9591 &0.7701 \\
Macro $Precision$   &0.9563 &0.7755 \\
Macro $Recall$ & 0.9629&0.7706\\
$Accuracy$ & 96.36\%&77.13\%\\
\hline
\end{tabular}
\end{table}

We also explored a two model hybrid approach to combat the challenges of classifying SH. In this approach, we cascade two probabilistic NNs with similar architectures except the output layers. The first model is the 8PNN to categorize plasma into CH, SR, SB and ICME and the second is a binary classification probabilistic NN to further categorize ICME values into MO and SH. The whole set up is hereafter called `cascaded 8PNN'. Data points classified as ICME in the first model of cascaded 8PNN are passed to the second model, which attempts to disambiguate the ICME into MO, SH, or `no prediction'. The confusion matrix for the second model of the cascaded 8PNN is shown in Figure \ref{fig5}(b) and the macro summary statistics are given in the Column 3 of the Table \ref{t3}. The features important in distinguishing MO from SH are shown in Figure \ref{fig5}(c). It is evident from Figure \ref{fig3} that a single NN cannot reliably distinguish between MO and SH. Creating two cascading models has practical benefits. Figure \ref{fig5}(a) shows that all four categories have high accuracy - false positives and false negatives are unlikely. Misclassifying MO as SH, and vice versa, in the second model is problematic, but we can still be highly confident we are observing an ICME. \par

\begin{figure}    
\centerline{\includegraphics[width=1.0\textwidth,clip=]{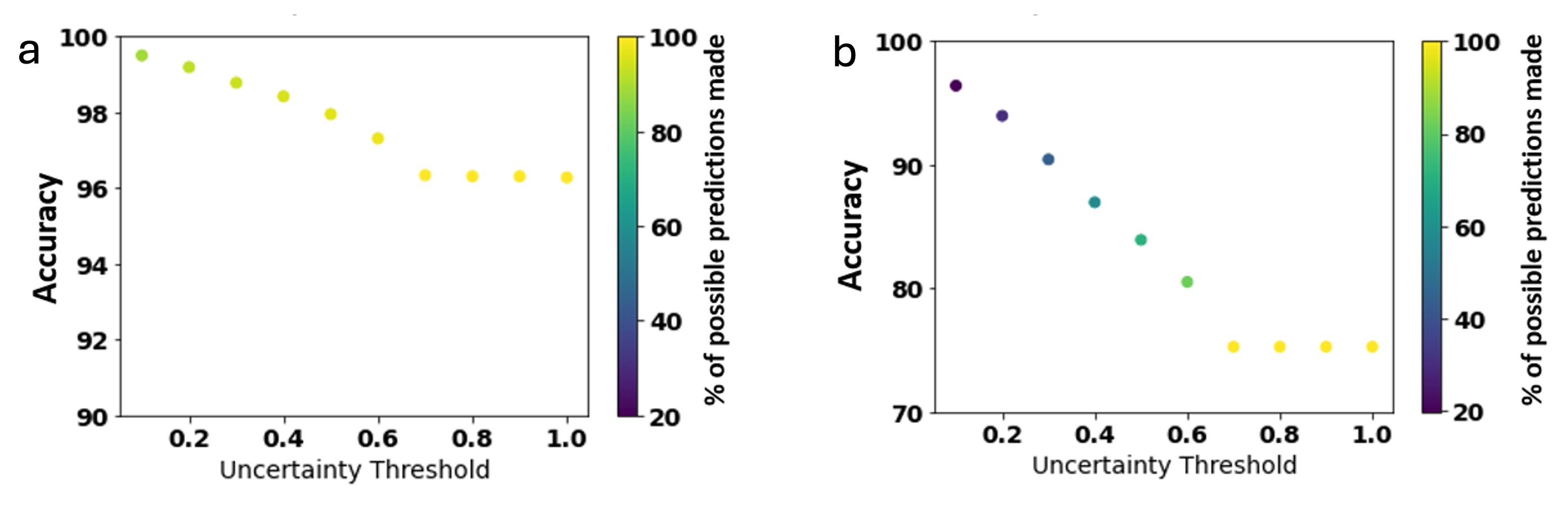}}
\caption{NN model $Accuracy$ as a function of user-specified uncertainty threshold for the (a) 8PNN model identifying CO, SB, SR and ICME and for the (b) binary classification NN model to classify identified ICME in SH and MO. 
}
\label{fig6}
\end{figure}

Returning to uncertainty thresholds, we explored how varying this threshold impacted the number of predictions made and the $Accuracy$ in both models of the cascaded 8PNN. The amount of allowed uncertainty is a subjective measure and our goal is to provide guidance for the user. Allowing more uncertainty leads to more predictions potentially resulting in higher misclassification. Figure \ref{fig6}(a) and (b) show the impact of varying the uncertainty threshold on the $Accuracy$ of the two models classifying CH, SB, SR, ICME and SH, MO, respectively.

\subsection{Application to Complex and Ejecta Events}

\cite{2018SoPh..293...25N} define `ejecta' (E) as solar wind transients without evident magnetic field rotation and those events with more than one magnetic field rotation as `complex' (Cx). This is different from \cite{2015JGRA..120...70X, 2017JGRA..12210910C} and \cite{2020ApJ...889..153R} who use the term `ejecta' to define ICME MCs. We applied our 8PNN model to all \cite{2018SoPh..293...25N} defined E and Cx events in the \href{https://wind.nasa.gov/ICME_catalog/ICME_catalog_viewer.php}{ICME Catalog}. We found that only 35\% of Cx and E were correctly classified as MOs. 19\% of Cx and E observations were incorrectly classified as CH, 14\% of them were incorrectly classified as SR, and 32\% of them were incorrectly classified as SB.\par

\begin{figure}    
\centerline{\includegraphics[width=1.0\textwidth,clip=]{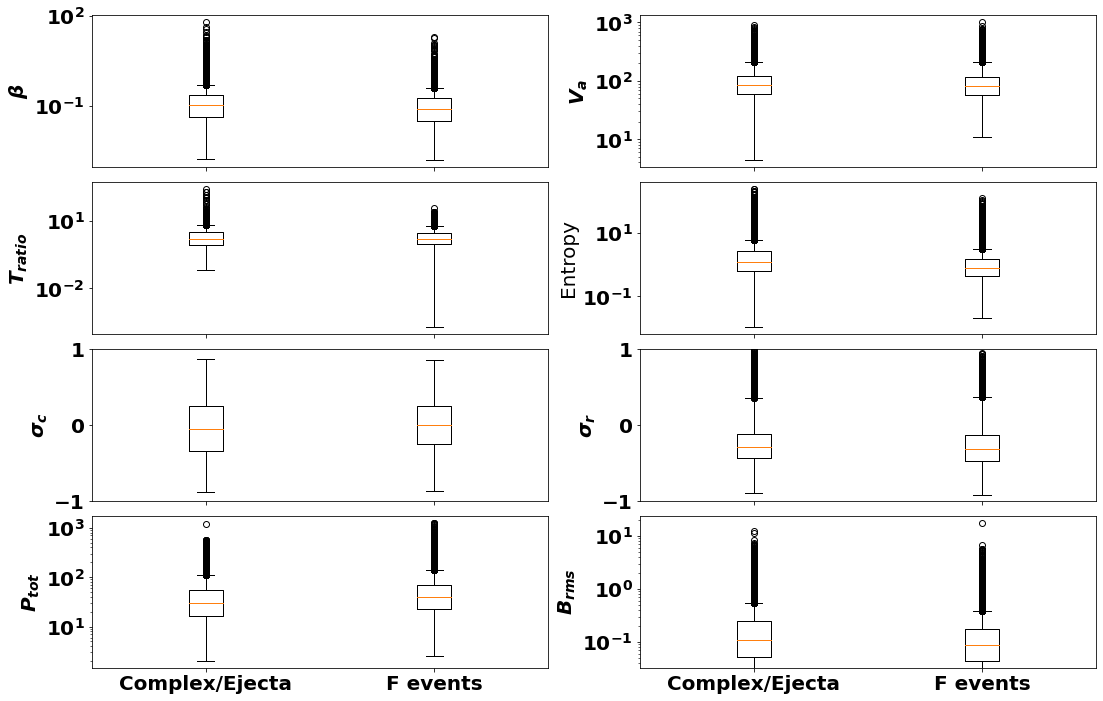}}
\caption{ Boxplots comparing Cx/E observations with Fr/F-/F+ observations based on the 8 input features -- plasma $\beta$, $V_a$, $T_{ratio}$, $Entropy$, $\sigma_c$, $\sigma_r$, $P_{tot}$ and $B_{rms}$.
}
\label{fig7}
\end{figure}

We then divided the Cx and E events into thirds and investigated where incorrect predictions were being made. We found that 65\% of the first third and middle third predictions were incorrect, and 64\% of the final third predictions were incorrect. Unlike Fr, F+, and F- events, Cx and E predictions were consistently incorrect across the duration of the MOs. To understand the difficulty of classifying these observations, we examined the characteristics of the Cx and E observations and found them to have more extreme values than Fr, F+, and F- observations. Comparison boxplots are shown in Figure \ref{fig7}. \par

The middle fifty percent of Cx/E observations and Fr/F-/F+ observations are similar across all eight parameters. However, the range of values seen and the extent of outliers varies considerably between Cx/E and Fr/F-/F+ observations. Unlike Fr, F+, F- events, which have high prediction accuracy, Cx and E events have more extreme plasma characteristics. Therefore, NNs trained on Fr, F+, and F- events do not transfer well to Cx and E events. This is interesting given that Fr/F-/F+, Cx, and E are all sub-types of MOs. The Cx and E are distinct enough, at least in terms of the 8 parameters studied here, that a single NN does not work on them all.\par

\par

\section{Summary and Conclusion}
\label{sec5}

This work combines solar wind characteristics with a probabilistic NN to identify different types of solar wind plasma depending on their origins. Following the study by \cite{2015JGRA..120...70X}, the NN was first trained with 3-parameter scheme -- $V_a$, $T_{ratio}$, and $Entropy$ -- to automatically classify solar wind plasma as CO, SB, SR or MO. We found that MOs are similar enough to MCs that the 3-parameter scheme can reliably detect this more general characterization of transients. The performance of the model increases when an additional five plasma parameters are included. These additional parameters -- $\sigma_c$, $\sigma_r$, $B_{rms}$, plasma $\beta$ and $P_{tot}$ -- were then used in training subsequent models. The contribution of more plasma parameters - namely plasma $\beta$ and $P_{tot}$ - increase the accuracy from $\sim88\%$ (in the case of 3PNN) to $\sim96\%$ (in the case of 8PNN). Despite $\sigma_c$ and $\sigma_r$ being proxies for the $O^{7+}/O^{6+}$ ratio, we find them of limited importance in our plasma type classifications. In 
\href{https://wind.nasa.gov/ICME_catalog/ICME_catalog_viewer.php}{ICME Catalog}, the MOs are categorised in three different types of flux ropes (Fr/F+/F-), ejecta (E) and complex (Cx). Although the catalog includes E and Cx as types of MOs, they have plasma characteristics distinct enough from Fr/F+/F- that both 3PNN and 8PNN trained on Fr/F+/F- perform poorly on E and Cx identifications.\par

Prior research on plasma state classification has not explored SH as a separate plasma state. The SH preceding the start of an MO is known to be highly variable, and sometimes not present at all. Our results demonstrate that the SH frequently has plasma characteristics similar to MOs. A clear distinction between the SH and MO is challenging at present for a machine learning model. 
In our study, we notice that the SH is most often misclassified as MO. Therefore, when the 8PNN model is used to classify plasma in five different categories (CO, SB, SR, MO and SH), 24\% (19\%) of SH (MO) cases are misclassified as MOs (SHs) and the model accuracy decreases to 76.94\%. This results as both regions tend to have higher $P_{tot}$, $T_{ratio}$, and $V_a$ while having lower $Entropy$ than the ambient solar wind. Our results show that a two step approach of first distinguishing ICMEs (a combined group of MO and SH) and then classifying ICME into MO and SH is more practically effective. We find that the 8PNN model has accuracy of 96.36\% in classifying plasma in CO, SH, SB and ICME (MO+SH) and 77.13\% in subsequently distinguishing ICME as SH or MO. Thus, a multi-model machine learning approach to space weather forecasting is likely to be the most beneficial.\par


Further, our work demonstrates that distinguishing amongst various plasma types is complex, due to variability in the characteristics of these states. This issue highlights the need for uncertainty estimation when classifying solar wind plasma events. To account for this uncertainty, we developed a probabilistic NN that provides both classification of these events and associated uncertainty of the classification. Because it is user-specified, the uncertainty threshold can accommodate different risk aversion levels. For example, a researcher who is risk-averse may select a lower uncertainty threshold. This will result in fewer events being classified. However, the classified events will have higher accuracy. This work focused on determination of aleatoric uncertainty, i.e. irreducible variability due to data. In future work we will examine epistemic uncertainty, i.e., uncertainty due to training a model for solar wind plasma classification. \par

Although this work provides a machine learning approach to automatically classifying solar wind plasma originating from different solar sources including coronal holes and streamer belts, we pay close attention to accurately identifying plasma from solar transients, namely ICMEs. ICMEs are one of the major drivers of space weather and research in accurately forecasting and identifying ICMEs has received significant attention. Our approaches contribute to auto-identification of ICMEs through categorising plasma using probabilistic NNs. We further attempted to classify ICMEs as SHs or MOs which are important segments of ICMEs frequently being observed to carry southward interplanetary magnetic field vector driving geomagnetic storms. SHs commonly precede MOs, meaning auto-identification of SHs may allow early forecasting of MOs and their prolonged and intense southward interplanetary magnetic field that drives geomagnetic storms. Our approach has the potential to be implemented on real-time solar wind data and support space weather now-casting and forecasting operations.\par

\begin{acks}
 We acknowledge NASA's Coordinated Data Analysis System (CDAS) and the open-source catalog of \href{https://wind.nasa.gov/ICME_catalog/ICME_catalog_viewer.php}{Wind ICMEs}. S.P is thankful for the support of NASA Postdoctoral Program fellowship. T.N-C. thanks the Solar Orbiter and Parker Solar Probe missions, Heliophysics Guest Investigator Grant 80NSSC23K0447 and the GSFC-Heliophysics Innovation Funds for support.
\end{acks}

\appendix 

\begin{table}[H]
\caption{ Summary of acronyms uses }
\label{appendixTable}
\begin{tabular}{p{0.2\linewidth} | p{0.6\linewidth}}    

\hline                    
Acronym & Full Name and Description\\
\hline
CH & Coronal-hole-origin plasma \\
CME & Coronal Mass Ejection \\
Cx & Complex event - a type of MO in which the ICME has a magnetic field configuration showing more than one rotation \\
E & Ejecta - a type of MO in which the ICME has a magnetic field configuration showing no evident rotation. We note that other authors use 'Ejecta' differently when referring to solar wind transients. \\
FR & Flux-rope - a type of MO in which the ICME has a magnetic field rotation suggestive of a flux-rope configuration \\
F+ & Large Rotation Flux-Rope - a type of MO in which the ICME has a magnetic field configuration containing a magnetic field rotation larger than a typical Flux-rope \\
F- & Small Rotation Flux-Rope - a type of MO in which the ICME has a magnetic field rotation smaller than a typical Flux-Rope \\
ICME & interplanetary ICME \\
MC & Magnetic Cloud - a subset of ICMEs containing an enhanced rotating magnetic field, low proton temperature, and low plasma beta \\
MO & Magnetic Obstacle - a less restrictive definition than Magnetic Cloud that allows for more general magnetic field configurations within ICMEs  \\
SB & streamer-belt-origin plasma \\
SH & sheath - compressed region that generally precedes solar wind transients and mostly form
during interplanetary propagation \\
SR & sector-reversal-region plasma, a sub-type of SB plasma \\
3PNN & 3 Parameter Neural Network - neural network that predicts solar wind type from 3 inputs \\
8PNN & 8 Parameter Neural Network - neural network that predicts solar wind type from 8 inputs \\
\hline
\end{tabular}
\end{table}

\begin{ethics}
    
\begin{conflict}
The authors declare that they have no conflicts of interest.
\end{conflict}

\begin{codeavailability}
    Code used to train our neural networks, along with the trained model weights, are publicly available in our \href{https://github.com/narock/solar_wind_plasma_type}{GitHub repository}.
\end{codeavailability}

\begin{dataavailability}
Wind and OMNI data were retrieved from NASA's Coordinated Data Analysis System (CDAS) using their Python-based library  \href{https://pypi.org/project/cdasws/}{cdasws 1.8.2.} The machine learning ready versions of the data used to train and test our models are available in the aforementioned GitHub repository. 
\end{dataavailability}

\end{ethics}

\bibliographystyle{spr-mp-sola}
\bibliography{sola_bibliography_example}

\end{document}